\documentclass[showpacs,preprintnumbers,amsmath,amssymb]{revtex4}

\usepackage{graphicx}
\usepackage{dcolumn}
\usepackage{bm}
\usepackage{url}
\usepackage{epsfig}
\usepackage{multirow}
\usepackage{tensor}
\usepackage{empheq}
\usepackage{amsmath}
\usepackage{color}
\definecolor{myblue}{rgb}{.8, .8, 1}

\def\be{\begin{equation}}
\def\ee{\end{equation}}
\def\ba{\begin{eqnarray}}
\def\ea{\end{eqnarray}}

\newcommand{\fr}[2]{\frac{#1}{#2}}
\def\kk{\vec{k}}

\def\D{\rm{D}}

\def\EdS{\rm{EdS}}

\def\ga{\mathrel{\raise.3ex\hbox{$>$\kern-.75em\lower1ex\hbox{$\sim$}}}}
\def\la{\mathrel{\raise.3ex\hbox{$<$\kern-.75em\lower1ex\hbox{$\sim$}}}}

\begin{document}


\leftline{KIAS-P14014}

\title{Exact Third-Order Density Perturbation and One-Loop Power Spectrum in General Dark Energy Models}


\author{Seokcheon Lee$^{1}$, Changbom Park$^{1}$, and Sang Gyu Biern$^{2}$}

\affiliation{
$^1$School of Physics, Korea Institute for Advanced Study, Heogiro 85, Seoul 130-722, Korea\\
$^2$Department of Physics and Astronomy, Seoul National University, Seoul 151-747, Korea}





\begin{abstract}
Under the standard perturbation theory (SPT), we obtain the fully consistent third-order density fluctuation and kernels for the general dark energy models without using the Einstein-de Sitter (EdS) universe assumption for the first time. We also show that even though the temporal and spatial components of the SPT solutions can not be separable, one can find the exact solutions to any order in general dark energy models. With these exact solutions, we obtain the less than \% error correction of one-loop matter power spectrum compared to that obtained from the EdS assumption for $k = 0.1 {\rm h\, Mpc}^{-1}$ mode at $z = 0$ (1, 1.5). Thus, the EdS assumption works very well at this scale. However, if one considers the correction for $P_{13}$, the error is about 6 (9, 11) \% for the same mode at $z = 0$ (1, 1.5). One absorbs $P_{13}$ into the linear power spectrum in the renormalized perturbation theory (RPT) and thus one should use the exact solution instead of the approximation one. The error on the resummed propagator $N$ of RPT is about 14 (8, 6) \% at $z =0$ (1, 1.5) for $k = 0.4 {\rm h\, Mpc}^{-1}$. For $k = 1 {\rm h\, Mpc}^{-1}$, the error correction of the total matter power spectrum is about 3.6 (4.6, 4.5) \% at $z = 0$ (1, 1.5). Upcoming observation is required to archive the sub-percent accuracy to provide the strong constraint on the dark energy and this consistent solution is prerequisite for the model comparison.

\end{abstract}



\maketitle

\setcounter{equation}{0}
The standard perturbation theory (SPT) has been widely used to investigate the correction to the linear power spectrum  in a quasi-nonlinear regime. The recent progress and the development of alternative analytical methods have been made \cite{0112551, 13112724}. The approximate recursion relations for the Fourier components of the $n$-th order matter density fluctuation $\hat{\delta}_{n}(\tau,\kk)$ and the divergence of the peculiar velocity $\hat{\theta}_{n}(\tau,\kk)$ has been obtained for the Einstein-de Sitter (EdS) universe \cite{Goroff, Makino}. When one extends the SPT to the general background universe, one uses the assumption that the dependence of the SPT solutions on the cosmological parameters is encoded in the linear growth factor, $D_{1}(a)$ \cite{0112551}. This is also confirmed for the dark energy models \cite{9807211, 08061437}. However, this argument is partly correct because one also needs to investigate the error on the power spectrum induced from EdS assumption ({\it i.e.} the value of the linear growth rate is equal to that of the square root of the matter energy density contrast, $f_1 \equiv \fr{d \ln D_{1}}{d \ln a} = \sqrt{\Omega_{m}}$ ). We obtain the exact kernels for $\hat{\delta}_{n}$ and $\hat{\theta}_{n}$ without using EdS assumption and study its effect on the power spectrum.

The renormalized perturbation theory (RPT) tries to reorganize the perturbative series expansion of SPT and resums some of the terms into a function that can be factorized out of the series \cite{0509418, 0509419}. This function is called as the resummed propagator and referred as $N$. All the kernels of the higher order power spectrum terms must be expressed as a product of kernels that correspond to full mode coupling terms and full propagator terms in order to make the resummation possible. If the kernels are approximated as a product of one-loop propagator kernels, then the resummed propagator is given by $N(k) \equiv \exp [P_{13}(k)/P_{{\rm lin}}(k)]$. We find that $P_{13}(k)$ using EdS assumption causes 6 $\sim$ 11 \% errors for $k = 0.5 {\rm h\, Mpc}^{-1}$ mode at $z = 0 \sim 1.5$ and these induce errors on $N$ about 11 $\sim$ 20 \%.

In addition to SPT, the Lagrangian perturbation theory (LPT) is an another widely used analytic technique for the quasi-linear perturbative expansion. There also have been studies to investigate the dark energy dependence on the linear growth factor in LPT \cite{12034260, 14012226}. Recently, we also obtain the kernels in the recursion relations without using EdS assumption in the LPT and investigate its consequences on the one-loop power spectrum \cite{14043813}.

In this {\em Letter}, we obtain the exact relations for the temporal and spatial components of the SPT solutions in general dark energy models up to third order. When we obtain the kernels, we remove the EdS assumption in the derivation and investigate the its effects on the observable quantities.

The equations of motion of $\hat{\delta}(\tau,\kk)$ and $\hat{\theta}(\tau,\kk)$ in the Fourier space are given by
\ba \fr{\partial \hat{\delta}}{\partial \tau} + \hat{\theta} &=& - \int d^3 k_{1} \int d^3 k_2 \delta_{\D}(\vec{k}_{12} - \vec{k}) \alpha(\vec{k}_1, \vec{k}_2) \hat{\theta} (\tau,\vec{k}_1) \hat{\delta} (\tau,\vec{k}_2) \label{massFT} \, , \\
\fr{\partial \hat{\theta}}{\partial \tau} + {\cal H} \hat{\theta} + \fr{3}{2} \Omega_{m} {\cal H}^2 \hat{\delta} &=& - \fr{1}{2} \int d^3 k_{1} \int d^3 k_2 \delta_{\D}(\vec{k}_{12} - \vec{k}) \beta(\vec{k}_1, \vec{k}_2) \hat{\theta} (\tau,\vec{k}_1) \hat{\theta} (\tau,\vec{k}_2) \label{EulerFT} \, ,
\ea
where $\tau$ is the conformal time, $\vec{k}_{12} \equiv \vec{k}_1 + \vec{k}_2$, $\delta_{\D}$ is the Dirac delta function, ${\cal H} \equiv \fr{1}{a} \fr{\partial a}{\partial \tau}$, $\Omega_{m}$ is the matter energy density contrast, $\alpha(\vec{k}_1, \vec{k}_2) \equiv \fr{\vec{k}_{12} \cdot \vec{k}_1}{k_1^2}$, and $\beta(\vec{k}_1, \vec{k}_2) \equiv \fr{k_{12}^2 (\vec{k}_1 \cdot \vec{k}_2)}{k_1^2 k_2^2}$.

Due to the mode coupling of the nonlinear terms shown in the right hand side of Eqs. (\ref{massFT}) - (\ref{EulerFT}), one needs to make a perturbative expansion in $\hat{\delta}$ and $\hat{\theta}$ \cite{0112551}. One can introduce the proper perturbative series of solutions for the fastest growing mode $D_{n}$
\ba \hat{\delta}(\tau,\vec{k}) &\equiv& \sum_{n=1}^{\infty} \hat{\delta}^{(n)} (\tau,\vec{k})  \label{hatdeltaS} \, , \\
\hat{\theta}(\tau,\vec{k}) &\equiv& \sum_{n=1}^{\infty} \hat{\theta}^{(n)} (\tau,\vec{k}) \label{hatthetaS} \, , \ea
where one can define the each order solution as
\ba \hat{\delta}^{(1)} (a,\vec{k}) &\equiv& D_{1}(a) \delta_{1}(\vec{k}) \, , \label{delta1} \\
\hat{\theta}^{(1)} (a,\vec{k}) &\equiv& D_{\theta 1}(a) \theta_{1}(\vec{k}) \equiv -a {\cal H} \fr{d D_1}{da} \delta_{1}(\vec{k})  \, , \label{theta1} \\
\hat{\delta}^{(2)} (a,\vec{k}) &\equiv& \sum_{i=1}^{2} D_{2i}(a) K_{2i}(\vec{k}) \equiv D_{1}^2(a) \sum_{i=1}^{2} c_{2i}(a) K_{2i}(\vec{k}) \, , \label{delta2} \\
\hat{\theta}^{(2)} (a,\vec{k}) &\equiv& \sum_{i=1}^{2} D_{\theta 2i}(a) K_{2i}(\vec{k}) \equiv a {\cal H} D_{1} \fr{d D_1}{da} \sum_{i=1}^{2} c_{\theta 2i}(a) K_{2i}(\vec{k}) \, , \label{theta2} \\
\hat{\delta}^{(3)} (a,\vec{k}) &\equiv& \sum_{i=1}^{6} D_{3i}(a) K_{3i}(\vec{k}) \equiv D_{1}^3(a) \sum_{i=1}^{6} c_{3i}(a) K_{3i}(\vec{k}) \, , \label{delta3} \\
\hat{\theta}^{(3)} (a,\vec{k}) &\equiv& \sum_{i=1}^{6} D_{\theta 3i}(a) K_{3i}(\vec{k}) \equiv a {\cal H} D_{1}^2 \fr{d D_1}{da} \sum_{i=1}^{6} c_{\theta 3i}(a) K_{3i}(\vec{k}) \, , \label{theta3} \ea
To be consistent with the current observation, we consider the dark energy dominated flat universe as a background model. It has been known that the $n$-th order fastest growing mode solutions are proportional to the $n$-th power of the linear growth factor $D_{1}$ ({\it i.e.} $D^{n} \propto D_{1}^n$) for the EdS universe. And this is not true for the general background models. There have been the investigations of the validity of these ansatz (\ref{hatdeltaS}) and (\ref{hatthetaS}) by using the different growth rates for $\hat{\delta}$ and $\hat{\theta}$ \cite{9807211, 08061437}. However, the improper decomposition of fastest mode solutions and the incorrect initial conditions are used for the $n$-th order growth rate in both cases (see Appendix).

If one takes a derivatives of Eq. (\ref{massFT}) and replace Eq. (\ref{EulerFT}) into it, then one obtains
\ba \fr{\partial^2 \hat{\delta}}{\partial \tau^2} + {\cal H} \fr{\partial \hat{\delta}}{\partial \tau} - \fr{3}{2} \Omega_{m} {\cal H}^2 \hat{\delta} &=& - {\cal H} \int d^3 k_{1} \int d^3 k_2 \delta_{\D}(\vec{k}_{12} - \vec{k}) \alpha(\vec{k}_1, \vec{k}_2) \hat{\theta} (\tau,\vec{k}_1) \hat{\delta} (\tau,\vec{k}_2) \nonumber \\ &&- \int d^3 k_{1} \int d^3 k_2 \delta_{\D}(\vec{k}_{12} - \vec{k}) \alpha(\vec{k}_1, \vec{k}_2) \Biggl[ \fr{\partial \hat{\theta} (\tau,\vec{k}_1)}{\partial \tau} \hat{\delta} (\tau,\vec{k}_2) + \hat{\theta} (\tau,\vec{k}_1) \fr{\partial \hat{\delta} (\tau,\vec{k}_2)}{\partial \tau} \Biggr] \nonumber \\ && + \fr{1}{2} \int d^3 k_{1} \int d^3 k_2 \delta_{\D}(\vec{k}_{12} - \vec{k}) \beta(\vec{k}_1, \vec{k}_2) \hat{\theta} (\tau,\vec{k}_1) \hat{\theta} (\tau,\vec{k}_2) \label{massEulerFT} \, ,
\ea
From the Eqs.(\ref{massFT}) and (\ref{massEulerFT}) , one obtains the expressions for the higher order solutions of $\hat{\delta}^{(2)}$, $\hat{\theta}^{(2)}$, and $\hat{\delta}^{(3)}$ as
\ba \hat{\delta}^{(2)}(a,\vec{k}) &\equiv& D_{21}(a) K_{21}(\vec{k}) + D_{22}(a) K_{22}(\vec{k}) \equiv D_{1}^2 \Biggl[ c_{21}(a) K_{21}(\vec{k}) + c_{22}(a) K_{22}(\vec{k}) \Biggr] \equiv D_{1}^2(a) \delta_{2}(a,\vec{k}) \nonumber \\
&\equiv& D_{1}^2 \int d^3 k_{1} \int d^3 k_2 \delta_{\D}(\vec{k}_{12} - \vec{k}) F_{2}^{(s)}(a, \vec{k}_1,\vec{k}_2) \delta_1(\vec{k}_1) \delta_1(\vec{k}_2) \, , \label{delta22} \\
\hat{\theta}^{(2)}(a,\vec{k}) &\equiv& D_{\theta 21}(a) K_{21}(\vec{k}) + D_{\theta 22}(a) K_{22}(\vec{k}) \equiv D_1 \fr{\partial D_{1}}{\partial \tau} \Biggl[ c_{\theta 21}(a) K_{21}(\vec{k}) + c_{\theta 22}(a) K_{22}(\vec{k}) \Biggr] \equiv  D_1 \fr{\partial D_{1}}{\partial \tau} \theta_{2}(a,\vec{k}) \nonumber \\
&\equiv& -D_1 \fr{\partial D_{1}}{\partial \tau} \int d^3 k_{1} \int d^3 k_2 \delta_{\D}(\vec{k}_{12} - \vec{k}) G_{2}^{(s)}(a,\vec{k}_1,\vec{k}_2) \delta_1(\vec{k}_1) \delta_1(\vec{k}_2) \, , \label{theta22} \\
\hat{\delta}^{(3)}(a,\vec{k}) &\equiv& D_{31}(a) K_{31}(\vec{k}) + \cdots +D_{36}(a) K_{36}(\vec{k}) \equiv D_1^3(a) \Biggl[ c_{31}(a) K_{31}(\vec{k}) + \cdots + c_{36}(a) K_{36}(\vec{k}) \Biggr] \nonumber \\ &\equiv& D_1^3(a) \int d^3 k_1 d^3 k_2 d^3 k_3 \delta_{\D}(\vec{k}_{123} - \vec{k}) F_{3}^{(s)}(a,\vec{k}_1,\vec{k}_2,\vec{k}_3) \delta_{1}(\vec{k}_1) \delta_{1}(\vec{k}_2) \delta_{1}(\vec{k}_3) \, , \label{delta32} \ea
where
\ba c_{2i} &=& \fr{D_{2i}}{D_1^2}\, , \,\,\,\, c_{\theta 2i} = \fr{D_{\theta 2i}}{D_1} \Bigl( \fr{\partial D_{1}}{\partial \tau} \Bigr)^{-1}\, , \,\,\,\, c_{3i} = \fr{D_{3i}}{D_1^3} \, , \label{ci} \\
K_{21}(\vec{k}) &=& -\int d^3 k_{1} \int d^3 k_2 \delta_{\D}(\vec{k}_{12} - \vec{k}) \alpha(\vec{k}_1, \vec{k}_2) \theta_1 (\vec{k}_1) \delta_1 (\vec{k}_2) \, , \label{K21} \\
K_{22}(\vec{k}) &=&  - \int d^3 k_{1} \int d^3 k_2 \delta_{\D}(\vec{k}_{12} - \vec{k}) \beta(\vec{k}_1, \vec{k}_2) \theta_1 (\vec{k}_1) \theta_1 (\vec{k}_2) \, , \label{K22} \\
F_{2}^{(s)}(a,\vec{k}_1,\vec{k}_2) &=& \fr{1}{2} \Biggl[ c_{21} \Bigl( \fr{\vec{k}_{12} \cdot \vec{k}_1}{k_1^2} + \fr{\vec{k}_{12} \cdot \vec{k}_2}{k_2^2} \Bigr) - 2 c_{22} \fr{k_{12}^2 (\vec{k}_1 \cdot \vec{k}_2)}{k_1^2 k_2^2} \Biggr] \label{F2s} \\
&=& c_{21} -2 c_{22} \Biggl(\fr{\vec{k}_1 \cdot \vec{k}_2}{k_1 k_2} \Biggr)^2 + \fr{1}{2} \Bigl(c_{21} -2 c_{22} \Bigr) \vec{k}_1 \cdot \vec{k}_2 \Biggl(\fr{1}{k_1^2} + \fr{1}{k_2^2} \Biggr) \, , \nonumber \\
G_{2}^{(s)}(a,\vec{k}_1,\vec{k}_2) &=& \fr{1}{2} \Biggl[ -c_{\theta 21} \Bigl( \fr{\vec{k}_{12} \cdot \vec{k}_1}{k_1^2} + \fr{\vec{k}_{12} \cdot \vec{k}_2}{k_2^2} \Bigr) + 2 c_{\theta 22} \fr{k_{12}^2 (\vec{k}_1 \cdot \vec{k}_2)}{k_1^2 k_2^2} \Biggr] \label{G2s} \\
&=& - c_{\theta 21} +2 c_{\theta 22} \Biggl(\fr{\vec{k}_1 \cdot \vec{k}_2}{k_1 k_2} \Biggr)^2 - \fr{1}{2} \Bigl(c_{\theta 21} -2 c_{\theta 22} \Bigr) \vec{k}_1 \cdot \vec{k}_2 \Biggl(\fr{1}{k_1^2} + \fr{1}{k_2^2} \Biggr) \, , \nonumber \\
F_{3}^{(s)}(a,\vec{k}_1,\vec{k}_2,\vec{k}_3) &=& \sum_{i=1}^{6} F_{3i}^{(s)}(a,\vec{k}_1,\vec{k}_2,\vec{k}_3) \, , \label{F3s}
\ea
where explicit forms of $F_{3i}^{(s)}$ are given in the appendix.

One can use the above equations to compute the power spectrum at any order in perturbation theory
\ba \langle \hat{\delta}(a,\vec{k}) \hat{\delta}(a,\vec{k}') \rangle &=&
D_{1}^2(a) \langle \delta_1(\vec{k}) \delta_1(\vec{k}') \rangle + D_1^4(a) \langle \delta_2(a, \vec{k}) \delta_2(a, \vec{k}') \rangle + D_1^4(a) \langle \delta_1(\vec{k}) \delta_3(a, \vec{k}') + \delta_3(a, \vec{k}) \delta_1(\vec{k}')  \rangle \nonumber \\ &\equiv& \Bigl( D_1^2 P_{11}(k) + D_1^4 P_{22}(a,k) + 2 D_1^4 P_{13}(a,k) + \cdots \Bigr) \delta_{\D}(\vec{k}+\vec{k}') \equiv P(a,k) \delta_{\D}(\vec{k}+\vec{k}')  \label{Pktau}  \ea
The one-loop power spectrum is defined as
\be P_{2}(a,k) = D_1^4(a) \Biggl[ P_{22}(a,k) + 2 P_{13}(a,k) \Biggr] \, , \label{P2} \ee where $P_{22}$ and $P_{13}$ are obtained as
\ba P_{22}(a,k) &=& 2 \int d^3 q P_{11}(q) P_{11} (|\vec{k}-\vec{q}\,|) \Bigl[ F_{2}^{(s)}(a,\vec{q}, \vec{k}-\vec{q}\,) \Bigr]^2 =
 \fr{(2 \pi)^{-2} k^3}{2} \int_{0}^{\infty} dr P_{11}(kr) \nonumber \\ &\times& \int_{-1}^{1} dx P_{11} \Bigl( k\sqrt{1+r^2-2rx} \Bigr) \Biggl[ \fr{(c_{21}+2c_{22}) r + (c_{21}-2c_{22}) x -2 c_{21} r x^2}{(1 + r^2 - 2rx)} \Biggr]^2 \, , \label{P22k} \\
2 P_{13}(k) &=& 6 P_{11}(k) \int d^3 q P_{11}(q) F_{3}^{(s)} (a, \vec{q}, \, -\vec{q}, \, \vec{k}\,) \label{P13k} \\
&=& (2\pi)^{-2} k^3 P_{11}(k) \int_{0}^{\infty} dr P_{11}(kr) \Biggl[ 2c_{35} r^{-2} -\fr{1}{3} \Bigl( 4 c_{31} -8 c_{32} +3c_{33} +24c_{35} - 16 c_{36} \Bigr) \nonumber \\ &-& \fr{1}{3}\Bigl(4 c_{31} -8c_{32} +12c_{33}-8c_{34}+6c_{35} \Bigr)r^2 + c_{33} r^4 + \Bigl(\fr{r^2-1}{r} \Bigr)^3 \ln \Bigl|\fr{1+r}{1-r} \Bigr| \Bigl(c_{35} - \fr{1}{2}c_{33}r^2 \Bigr) \Biggr] \nonumber  \, ,\ea
where $r = \fr{q}{k}$ and $x = \fr{\vec{q} \cdot \vec{k}}{q k}$. The above equations (\ref{P22k}) and (\ref{P13k}) are identical to Eqs. (2.24)
and (2.25) of \cite{Makino} when one replace the coefficients of higher solutions $c_{2i}$ and $c_{3i}$ with those given in Eqs.(\ref{c2iEdS}) and (\ref{c3iEdS}). Thus, the terms with $c_{2i}$ and $c_{3i}$ represent the dark energy effect on the one-loop power spectrum.
Now we obtain the one-loop power spectrum for $\Lambda$CDM model. We run the camb to obtain the linear power spectrum \cite{camb} using $\Omega_{b0} = 0.044$, $\Omega_{m0} = 0.26$, $h = 0.72$, $n_{s} = 0.96$, and the numerical integration range for $q$ in Eqs. (\ref{P22k}) and (\ref{P13k}) is $10^{-6} \leq q \leq 10^{2}$. 

\begin{center}
\begin{figure}
\vspace{1.5cm}
\centerline{\psfig{file=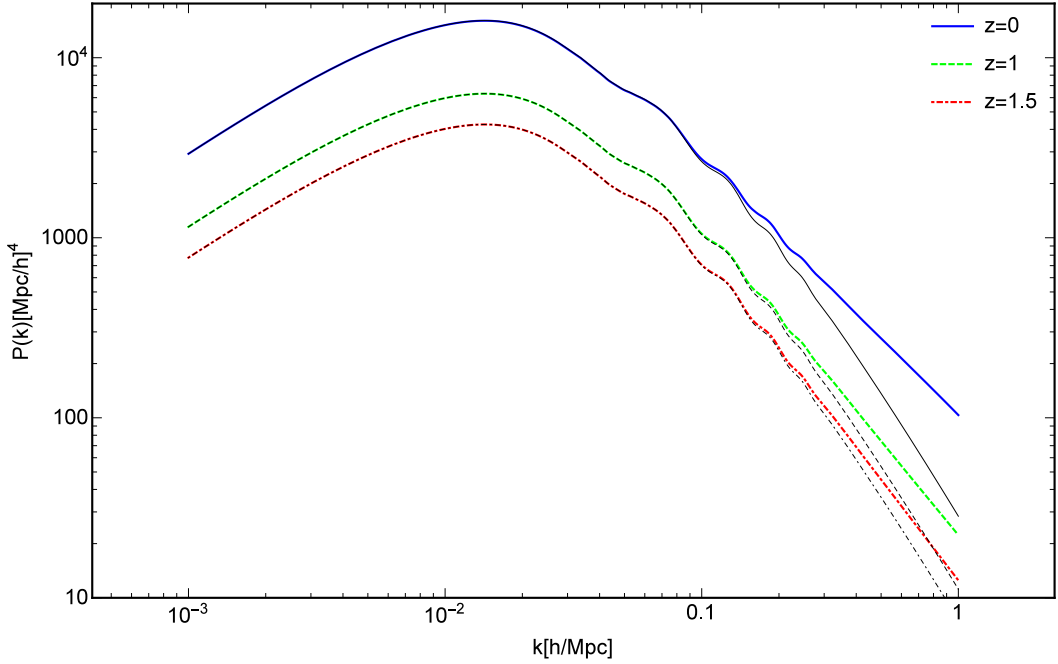, width=14cm}}
\vspace{-0.1cm}
\caption{Both the linear matter power spectra (thin lines) and the nonlinear matter power spectra with one-loop correction (thick lines) at $z = 0, 1.0,$ and 1.5 (solid, dotted, and dotdashed lines) for $\Omega_{m0} = 0.26$ $\Lambda$CDM model.} \label{fig1}
\end{figure}
\end{center}

In Fig. \ref{fig1}, we show both the linear power spectra $P_{\rm{L}}$ (thin lines) and the nonlinear power spectra $P_{\rm{NL}} = P_{\rm{L}} + P_{2}$ (thick lines) at the different redshift $z =$ 0 (solid), 1.0 (dotted), and 1.5 (dotdashed), respectively. We demonstrate the $\Lambda$CDM model with $\Omega_{m0} = 0.26$ in this figure. As one expects, the nonlinear power spectra are not simply enhanced by multiplying the differences of the square of the growth factor $D_1^2$ at the different redshifts. One also needs to emphasize that the exact kernels Eqs.(\ref{F2s}), (\ref{G2s}) and (\ref{F31s})-(\ref{F36s}) also depend on time. The coefficient of each kernel changes at the different observational epoch.

\begin{figure}
\centering
\vspace{1.5cm}
\begin{tabular}{cc}
\epsfig{file=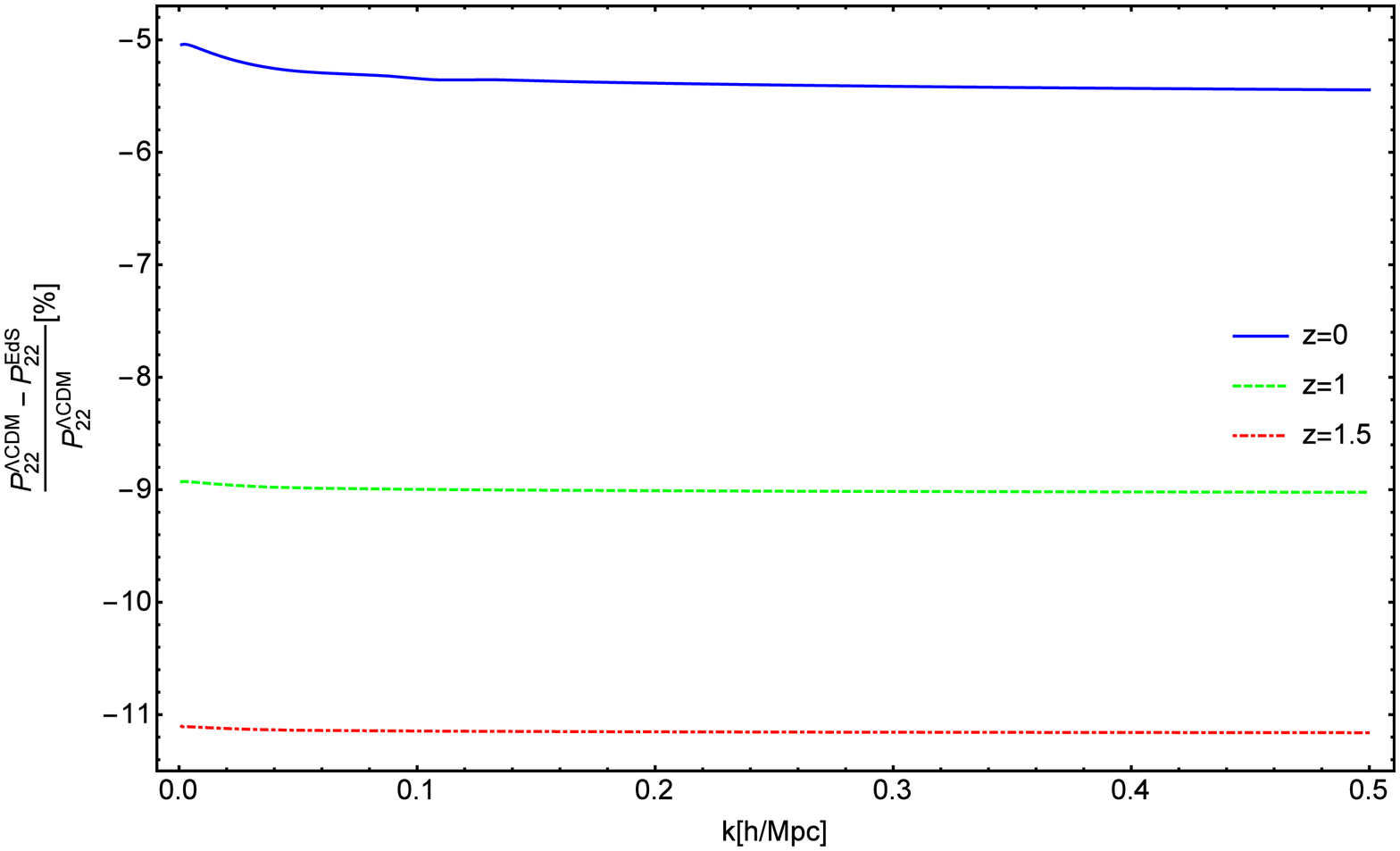,width=0.5\linewidth,clip=} &
\epsfig{file=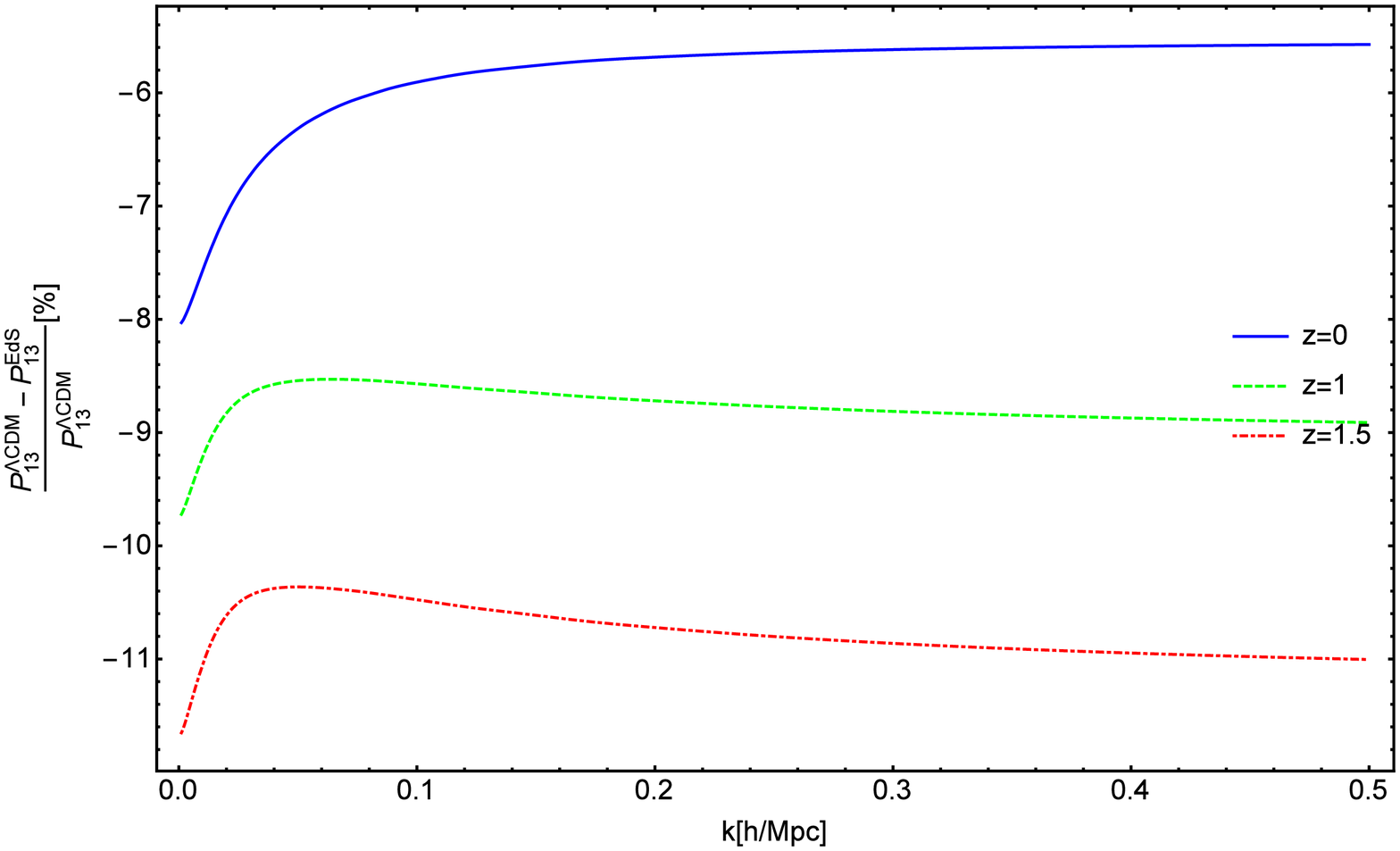,width=0.5\linewidth,clip=} \\
\end{tabular}
\vspace{-0.5cm}
\caption{Errors in $P_{22}$ and $P_{13}$ a) Differences between the correct $P_{22}$ and the one with EdS assumption at the different epoches. The solid, dashed, and dotdashed lines correspond to $z = 0$, 1.0, and 1.5, respectively. a) Differences between the correct $P_{13}$ and the EdS assumed $P_{13}$ at different epoches.} \label{fig2}
\end{figure}

Now, we investigate the corrections in $P_{22}$ and $P_{13}$ compared to those using the EdS assumption. 
As one expects, the effect of the removing EdS assumption on $P_{22}$ and $P_{13}$ becomes larger as $z$ increases. 
This is due to the fact that we use the Gaussianity initial conditions for the perturbed quantities. The coefficients $c_{21}$-$c_{36}$ approach to those of $EdS$ models as $z$ decreases. This causes the fact that the kernels based on the EdS assumption deviate from the exact ones as $z$ decreases. Thus, the exact $P_{22}$ and $P_{13}$ show the larger deviations from the EdS assumed $P{22}$ and $P_{13}$ as $z$ increases. This is shown in Fig. \ref{fig2}. $P_{22}^{\rm{\Lambda CDM}}$ and $P_{13}^{\rm{\Lambda CDM}}$ mean the exact one loop corrections based on the $\Lambda$CDM models using the exact solution. While $P_{22}^{\rm{EdS}}$ and $P_{13}^{\rm{EdS}}$ mean the one loop corrections based on the EdS assumed kernels. In the left panel of Fig. \ref{fig2}, we show the errors in $P_{22}$ at the different redshift. The solid, dotted, and dotdashed lines correspond to errors of $P_{22}$ at $z = 0$, 1.0, and 1.5, respectively. The differences are about 5 (9, 11) \% for $k = 0.1 {\rm h\, Mpc}^{-1}$ mode at $z =$ 0 (1.0, 1.5).  In the right panel of Fig. \ref{fig2}, we show the errors in $P_{13}$ at the different redshift. We use the same notation as the left panel. The differences between the exact and EdS assumed $P_{13}$ are about 6 (9, 11) \% for $k = 0.1 {\rm h\, Mpc}^{-1}$ mode at $z =$ 0 (1.0, 1.5).

\begin{figure}
\centering
\vspace{1.5cm}
\begin{tabular}{cc}
\epsfig{file=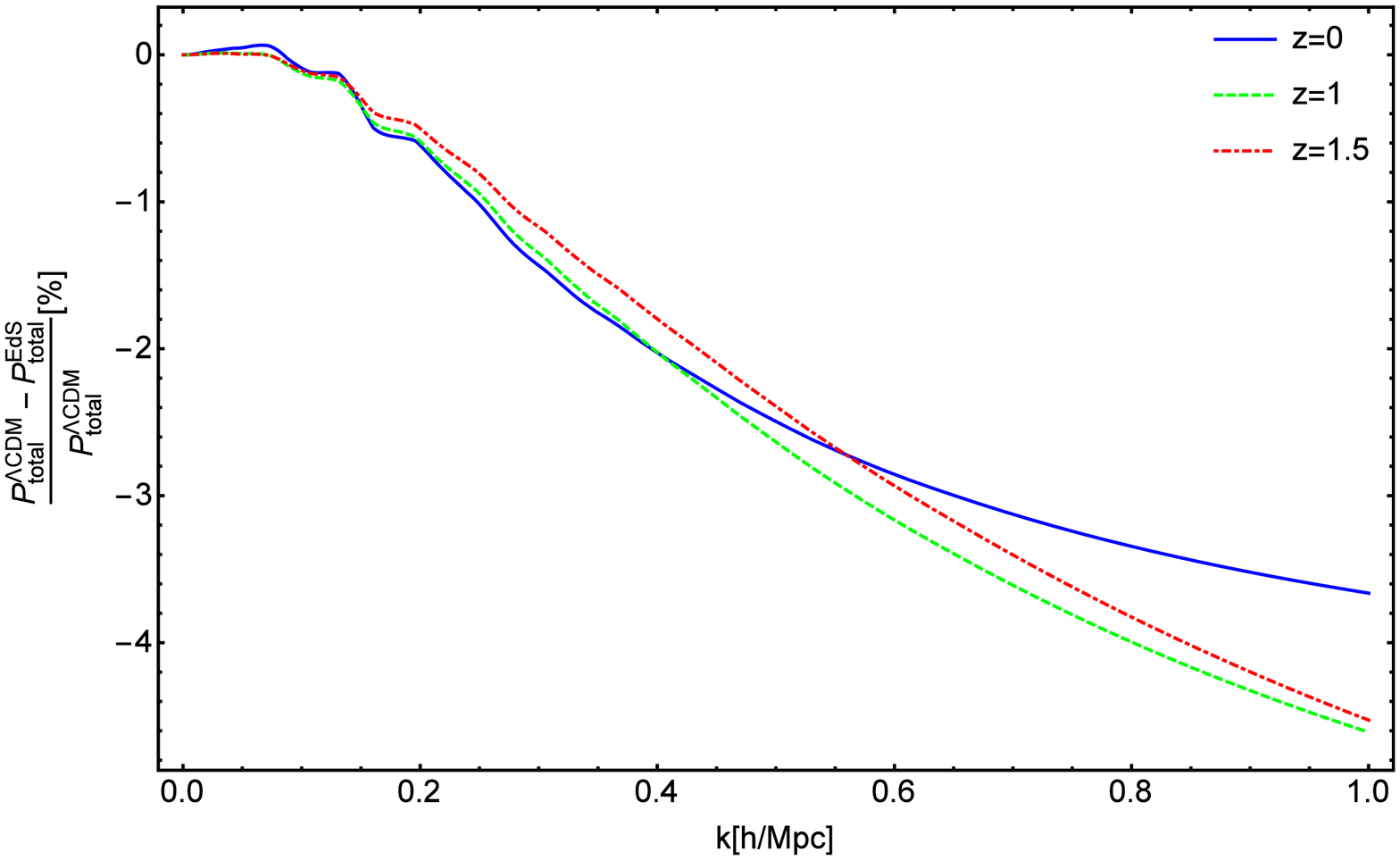,width=0.5\linewidth,clip=} &
\epsfig{file=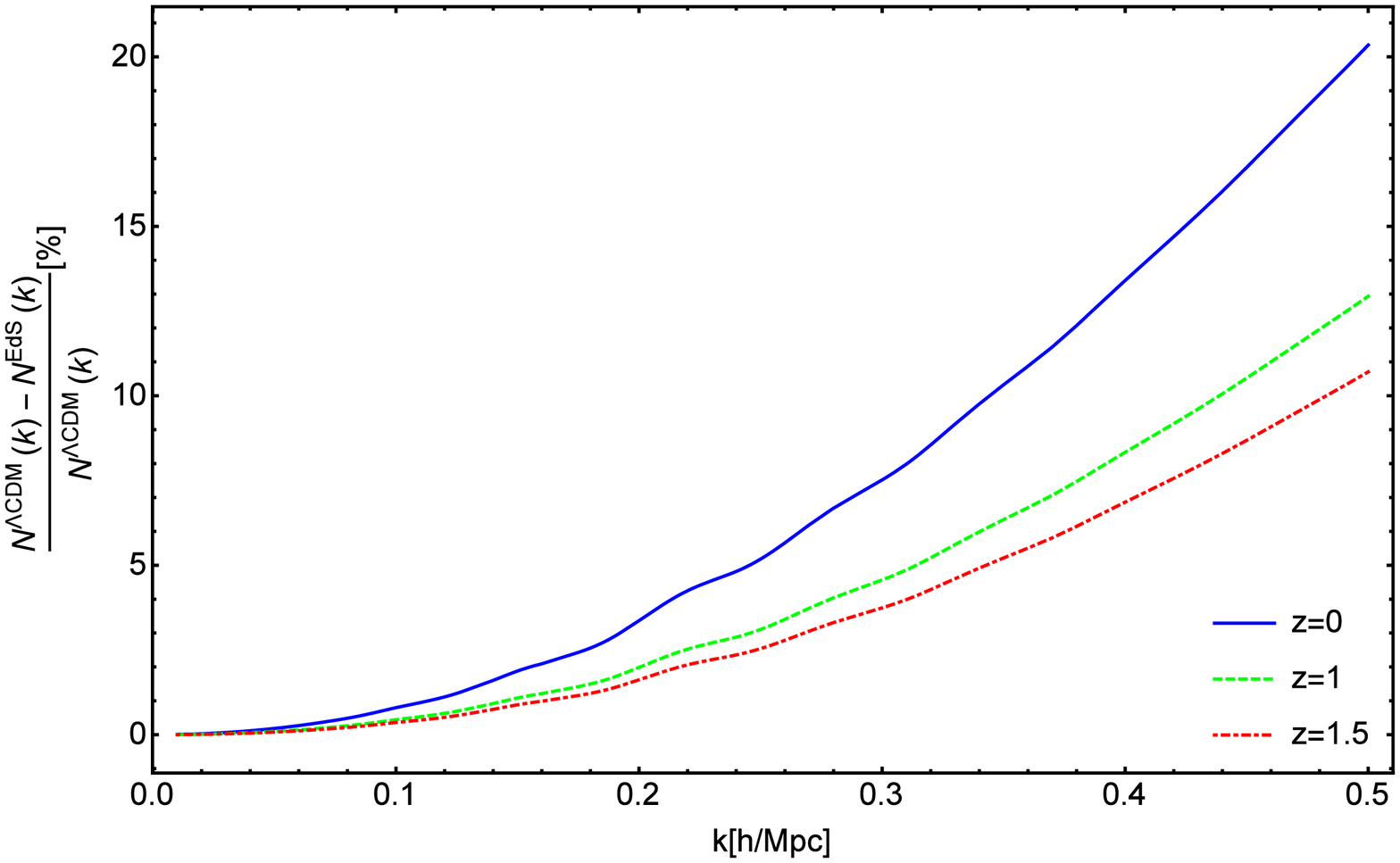,width=0.5\linewidth,clip=} \\
\end{tabular}
\vspace{-0.5cm}
\caption{Errors in $P_{\rm{total}}$ and $N$ a) Differences between the correct $P_{\rm{total}}$ and the one with $\lambda =1$ (EdS) assumption at the different epoches. The solid, dashed, and dotdashed lines correspond to $z = 0$, 0.5, and 1, respectively. a) Differences between the correct $P_{NL}$ and the $\lambda =1$ (EdS) assumed $P_{NL}$ at different epoches.} \label{fig3}
\end{figure}

We show the corrections on $P_{\rm{NL}} \equiv P_{\rm{total}}$ and the resummed propagator $N$.
The one loop correction is sum of the $P_{22}$ and $P_{13}$. However, $P_{22}$ and $P_{13}$ have the different signs. Thus, if one considers the nonlinear power spectrum with the one loop correction, then the correction due to using the exact solution is very small compared to the nonlinear power spectrum based on EdS assumption. $P_{\rm{NL}} = P_{\rm{L}} + P_{2}$ where $P_{2} = P_{22} + P_{13}$. As we show in the Fig. \ref{fig2}, each correction at each mode is about same at any epoch. Thus, the corrections on $P_{2}$ are canceled each other. This is shown in the left panel of Fig. \ref{fig3}. $P_{\rm{total}}^{\rm{\Lambda CDM}}$ means the exact nonlinear matter power spectrum based on the $\Lambda$CDM models using the exact solution. While $P_{\rm{total}}^{\rm{EdS}}$ means the nonlinear matter power spectrum based on the EdS assumption. The solid, dotted, and dotdashed lines correspond to errors of $P_{\rm{total}}$ at $z = 0$, 1.0, and 1.5, respectively. The present nonlinear matter power spectrum is dominated by the one loop power spectrum at small scale $k \geq 0.1$. The correction for the total matter power spectrum is about 2 \% for $k = 0.4$ h/Mpc at any epoch. Thus, the EdS assumed nonlinear power spectrum is not a bad approximation. However, if one expands the SPT into RPT, then one needs to use the exact solution. This is shown in the right panel of Fig. \ref{fig3} by using the resummed propagator $N$. For the same mode, the deviations of $N$ from the exact values are about 14 (8, 6) \% at $z =$ 0 (1.0, 1.5). Thus, if one uses the EdS assumed nonlinear $P_{13}$, then one is not able to avoid these amount of errors on the $N$.

The upcoming redshift surveys of galaxies such as BOSS, eBOSS, PFS, EUCLID, and MS-DESI will provide observational data of large scale structure
of the universe in larger volume with higher density. The analysis of these observational data requires very accurate theoretical modeling down to the quasi-linear regime. In this {\em Letter}, we present an accurate perturbation theory without adopting the EdS assumption. The obtained results are general for any background universe model including time varying dark energy models, and will be useful for studies of future surveys.

\section*{Acknowledgments}
This work were carried out using computing resources of KIAS Center for Advanced Computation. S.L would like to thank for the hospitality at APCTP during the program TRP.

\renewcommand{\theequation}{A-\arabic{equation}}
\setcounter{equation}{0}  
\section*{APPENDIX}  

In this section, we show the spatial and temporal solutions of the each order by using Eqs. (\ref{EulerFT}) and (\ref{massEulerFT}).
The equations for the first order solution of $\delta^{(1)}(\tau,\kk)$ and $\theta^{(1)}(\tau,\kk)$ are given by
\ba \Biggl[ \fr{\partial^2 D_{1}}{\partial \tau^2} + {\cal H} \fr{\partial D_{1}}{\partial \tau} - \fr{3}{2} \Omega_{m} {\cal H}^2 D_{1} \Biggr] \delta_{1}(\vec{k}) &=& 0 \, , \label{delta12} \\
D_{\theta 1} \theta_{1}(\kk) &=& - \fr{\partial D_1}{\partial \tau} \delta_{1}(\kk) \, . \label{theta12} \ea
From the above Eqs. (\ref{delta12}) and (\ref{theta12}), one obtains
\ba && \fr{d^2 D_{1}}{da^2} + \fr{3}{2a} \Bigl(1 - w \Omega_{de} \Bigr) \fr{d D_{1}}{da} - \fr{3}{2a^2} \Omega_{m} D_{1} = 0 \, , \label{D1eq} \\
&& D_{\theta 1}(a) = a {\cal H} \fr{\partial D_1}{\partial a}, \,\,\,\,
\theta_{1}(\kk) = - \delta_{1}(\kk) \, , \label{theta1k} \ea
where we use
\ba \fr{d D_1}{d \tau} &=& a {\cal H} \fr{d D_1}{da}, \,\,\,\,
\fr{d^2 D_1}{d \tau^2} = (a {\cal H})^2 \fr{d^2 D_1}{d a^2} + \Bigl( a {\cal H}^2 + a \fr{d {\cal H}}{d \tau} \Bigr) \fr{d D_1}{d a} \label{D1tau2} \, , \\
{\cal H} &=& a H, \,\,\,
\fr{d {\cal H}}{d \tau} = {\cal H}^2 + a^2 \fr{d H}{d t}, \,\,\,
\fr{1}{{\cal H}^2} \fr{d{\cal H}}{d \tau} = 1 + \fr{1}{H^2} \fr{d H}{d t} = 1 -\fr{3}{2} \Bigl( 1 + w \Omega_{de} \Bigr) \label{calHtau2} \, . \ea
If one uses the fact that the dark energy is dominated only at the late universe, then one can adopt the EdS conditions ({\it i.e.} $\Omega_{m} = 1$) for $D_1$ at early time ({it i.e.} $a_i$),
\be D_{1}(a_i) = a_i\, , \,\,\,\,\, {\rm and} \,\,\,\, \fr{d D_1}{da} \Bigl|_{a=a_i} = 1 \, . \label{D1EdS} \ee
Thus, one can obtain the exact solution for $D_1(a)$ for any dark energy model from Eqs.(\ref{delta12}) and (\ref{D1EdS}) except for the early dark energy one \cite{9408025,0012139,0504650}.

One can repeat the same process for $\hat{\delta}^{(2)}(\tau,\vec{k})$ to get
\ba \Biggl[ \fr{\partial^2 D_{21}}{\partial \tau^2} + {\cal H} \fr{\partial D_{21}}{\partial \tau} - \fr{3}{2} \Omega_{m} {\cal H}^2 D_{21} \Biggr] K_{21}(\vec{k}) &=& \Biggl[{\cal H} D_{1} \fr{\partial D_{1}}{\partial \tau} + D_1 \fr{\partial^2 D_{1}}{\partial \tau^2} + \Bigl(\fr{\partial D_1}{\partial \tau} \Bigr)^2 \Biggr] \label{delta21} \\
&\times& \Biggl[ - \int d^3 k_{1} \int d^3 k_2 \delta_{\D}(\vec{k}_{12} - \vec{k}) \alpha(\vec{k}_1, \vec{k}_2) \theta_1 (\vec{k}_1) \delta_1 (\vec{k}_2) \Biggr] \nonumber \, , \\
\Biggl[ \fr{\partial^2 D_{22}}{\partial \tau^2} + {\cal H} \fr{\partial D_{22}}{\partial \tau} - \fr{3}{2} \Omega_{m} {\cal H}^2 D_{22} \Biggr] K_{22}(\vec{k}) &=& -\fr{1}{2} \Bigl(\fr{\partial D_1}{\partial \tau} \Bigr)^2 \label{delta22} \\
&\times&  \Biggl[ - \int d^3 k_{1} \int d^3 k_2 \delta_{\D}(\vec{k}_{12} - \vec{k}) \beta(\vec{k}_1, \vec{k}_2) \theta_1 (\vec{k}_1) \theta_1 (\vec{k}_2) \Biggr] \nonumber \, .  \ea
If we adopt the initial zero non-Gaussianity of the higher order solutions ($\delta^{(n)} = 0$),  then one can obtain the equations for the fastest growing mode solutions with the initial Gaussianity and the EdS initial conditions
\ba \fr{d^2 D_{21}}{da^2} + \fr{3}{2a} \Bigl(1 - w \Omega_{de} \Bigr) \fr{d D_{21}}{da} - \fr{3}{2a^2} \Omega_{m} D_{21} &=& \fr{3}{2a^2} \Omega_{m} D_1^2 + \Bigl(\fr{d D_1}{da} \Bigr)^2 \label{D21eq} \\
&{\rm with}& D_{21}(a_i) = 0, \hspace{0.1in} \fr{d D_{21}}{da} \Bigl|_{a_i} = \fr{5}{7} a_i \nonumber \, , \\
\fr{d^2 D_{22}}{da^2} + \fr{3}{2a} \Bigl(1 - w \Omega_{de} \Bigr) \fr{d D_{22}}{da} - \fr{3}{2a^2} \Omega_{m} D_{22} &=& -\fr{1}{2} \Bigl(\fr{d D_1}{da} \Bigr)^2 \label{D22eq} \\
&{\rm with}& D_{22}(a_i) = 0, \hspace{0.1in} \fr{d D_{22}}{da} \Bigl|_{a_i} = -\fr{1}{7} a_i \nonumber \, ,\ea
where we use the fastest growing mode solutions for the EdS universe
\be D_{21}^{(\EdS)} =  \fr{5}{7} a^2 - \fr{5}{7} a_i a, \,\,\,\,
D_{22}^{(\EdS)} = -\fr{1}{7} a^2 + \fr{1}{7} a_i a \, . \label{D22EdS} \ee
Often it is knows as the EdS coefficient as
\be c_{21} = \fr{5}{7}\, , \,\,\,\, c_{22} = - \fr{1}{7} \, . \label{c2iEdS} \ee
However, this is not the coefficients for the fastest growing mode solutions because of the existence of the second terms in Eq.(\ref{D22EdS}).
From Eq.(\ref{EulerFT}), one can obtain equations for $\hat{\theta}^{(2)}$ by using other solutions
\ba D_{\theta 21}(a) K_{21}(\kk) + D_{\theta 22}(a) K_{22}(\kk) &=& a {\cal H} \Bigl[-\fr{d D_{21}}{d a} + D_1 \fr{d D_1}{d a} \Bigr] K_{21}(\kk) - a {\cal H} \fr{d D_{22}}{d a} K_{22}(\kk) \nonumber \, , \\
&\equiv& a {\cal H} D_{1} \fr{d D_{1}}{d a} \Bigl[ c_{\theta 21} K_{21}(\kk) + c_{\theta 22} K_{22}(\kk) \Bigr] \, . \label{theta22} \ea
From Eqs.(\ref{delta22})-(\ref{theta22})
\ba \hat{\delta}^{(2)}(a,\vec{k}) &\equiv& D_{21}(a) K_{21}(\vec{k}) + D_{22}(a) K_{22}(\vec{k}) \equiv D_{1}^2 \Biggl[ c_{21} K_{21}(\vec{k}) + c_{22} K_{22}(\vec{k}) \Biggr] \equiv D_{1}^2(a) \delta_{2}(a,\vec{k}) \nonumber \\
&=& D_{1}^2 \int d^3 k_{1} d^3 k_2 \delta_{\D}(\vec{k}_{12} - \vec{k}) \Bigl[ c_{21}(a) \alpha(\vec{k}_1, \vec{k}_2) - c_{22}(a) \beta(\vec{k}_1, \vec{k}_2) \Bigr] \delta_1(\vec{k}_1) \delta_1(\vec{k}_2) \nonumber \\
&\equiv& D_{1}^2 \int d^3 k_{1} d^3 k_2 \delta_{\D}(\vec{k}_{12} - \vec{k}) F_{2}(a, \vec{k}_1,\vec{k}_2) \delta_1(\vec{k}_1) \delta_1(\vec{k}_2) \label{delta23} \\
&\equiv& D_{1}^2 \int d^3 k_{1} \int d^3 k_2 \delta_{\D}(\vec{k}_{12} - \vec{k}) F_{2}^{(s)}(a, \vec{k}_1,\vec{k}_2) \delta_1(\vec{k}_1) \delta_1(\vec{k}_2) \nonumber \, , \\
F_{2}^{(s)}(a,\vec{k}_1,\vec{k}_2) &=& \fr{1}{2} \Biggl[ c_{21} \Bigl( \fr{\vec{k}_{12} \cdot \vec{k}_1}{k_1^2} + \fr{\vec{k}_{12} \cdot \vec{k}_2}{k_2^2} \Bigr) - 2 c_{22} \fr{k_{12}^2 (\vec{k}_1 \cdot \vec{k}_2)}{k_1^2 k_2^2} \Biggr] \nonumber \\
&=& c_{21} -2 c_{22} \Biggl(\fr{\vec{k}_1 \cdot \vec{k}_2}{k_1 k_2} \Biggr)^2 + \fr{1}{2} \Bigl(c_{21} -2 c_{22} \Bigr) \vec{k}_1 \cdot \vec{k}_2 \Biggl(\fr{1}{k_1^2} + \fr{1}{k_2^2} \Biggr) \label{F2s} \, , \\
\hat{\theta}^{(2)}(a,\vec{k}) &\equiv& D_{21}(a) K_{21}(\vec{k}) + D_{22}(a) K_{22}(\vec{k}) \equiv D_1 \fr{\partial D_{1}}{\partial \tau} \Biggl[ c_{\theta 21}(a) K_{21}(\vec{k}) + c_{\theta 22}(a) K_{22}(\vec{k}) \Biggr] \nonumber \\
&=& a {\cal H} \Bigl[-\fr{d D_{21}}{d a} + D_1 \fr{d D_{1}}{d a} \Bigr] K_{21}(\vec{k}) - a {\cal H} \fr{d D_{22}}{d a} K_{22}(\vec{k}) \equiv a {\cal H} D_1 \fr{d D_{1}}{d a} \theta_{2}(a,\vec{k})  \nonumber \\
&=& -a {\cal H} D_1 \fr{d D_{1}}{d a} \int d^3 k_{1} \int d^3 k_2 \delta_{\D}(\vec{k}_{12} - \vec{k}) \Bigl[ -c_{\theta 21}(a) \alpha(\vec{k}_1, \vec{k}_2) + c_{\theta 22}(a) \beta(\vec{k}_1, \vec{k}_2) \Bigr] \delta_1(\vec{k}_1) \delta_1(\vec{k}_2) \nonumber \\
&\equiv& -a {\cal H} D_1 \fr{d D_{1}}{d a} \int d^3 k_{1} \int d^3 k_2 \delta_{\D}(\vec{k}_{12} - \vec{k}) G_{2}(a,\vec{k}_1,\vec{k}_2) \delta_1(\vec{k}_1) \delta_1(\vec{k}_2) \label{theta23} \\
&\equiv& -a {\cal H} D_1 \fr{d D_{1}}{d a} \int d^3 k_{1} \int d^3 k_2 \delta_{\D}(\vec{k}_{12} - \vec{k}) G_{2}^{(s)}(a,\vec{k}_1,\vec{k}_2) \delta_1(\vec{k}_1) \delta_1(\vec{k}_2) \nonumber \, , \\
G_{2}^{(s)}(a,\vec{k}_1,\vec{k}_2) &=& \fr{1}{2} \Biggl[ -c_{\theta 21} \Bigl( \fr{\vec{k}_{12} \cdot \vec{k}_1}{k_1^2} + \fr{\vec{k}_{12} \cdot \vec{k}_2}{k_2^2} \Bigr) + 2 c_{\theta 22} \fr{k_{12}^2 (\vec{k}_1 \cdot \vec{k}_2)}{k_1^2 k_2^2} \Biggr] \nonumber \\
&=& - c_{\theta 21} +2 c_{\theta 22} \Biggl(\fr{\vec{k}_1 \cdot \vec{k}_2}{k_1 k_2} \Biggr)^2 - \fr{1}{2} \Bigl(c_{\theta 21} -2 c_{\theta 22} \Bigr) \vec{k}_1 \cdot \vec{k}_2 \Biggl(\fr{1}{k_1^2} + \fr{1}{k_2^2} \Biggr) \label{G2s} \, .
\ea

Now one can obtain the third order solutions from the previous solutions up to the second order. One can write the third order solution
\ba \hat{\delta}^{(3)}(a,\vec{k}) &\equiv& \sum_{i=1}^{6} D_{3i}(a) K_{3i}(\vec{k}) = D_{31}(a) K_{31}(\vec{k}) + \cdots +D_{36}(a) K_{36}(\vec{k}) \label{delta3}\\ &\equiv& D_1^3(a) \Biggl[ c_{31}(a) K_{31}(\vec{k}) + \cdots + c_{36}(a) K_{36}(\vec{k}) \Biggr] \nonumber \\ &\equiv& D_1^3(a) \int d^3 k_1 d^3 k_2 d^3 k_3 \delta_{\D}(\vec{k}_{123} - \vec{k}) F_{3}(a,\vec{k}_1,\vec{k}_2,\vec{k}_3) \delta_{1}(\vec{k}_1) \delta_{1}(\vec{k}_2) \delta_{1}(\vec{k}_3) \nonumber \\
&\equiv& D_1^3(a) \int d^3 k_1 d^3 k_2 d^3 k_3 \delta_{\D}(\vec{k}_{123} - \vec{k}) F_{3}^{(s)}(a,\vec{k}_1,\vec{k}_2,\vec{k}_3) \delta_{1}(\vec{k}_1) \delta_{1}(\vec{k}_2) \delta_{1}(\vec{k}_3) \, . \nonumber \ea
If one replaces Eq.(\ref{delta3}) into Eq.(\ref{massEulerFT}), then one obtains
\ba \Biggl[ \fr{\partial^2 D_{31}}{\partial \tau^2} + {\cal H} \fr{\partial D_{31}}{\partial \tau} - \fr{3}{2} \Omega_{m} {\cal H}^2 D_{31} \Biggr] K_{31}(\vec{k}) &=& \Biggl[\Bigl(\fr{\partial^2 D_{1}}{\partial \tau^2} + {\cal H} \fr{\partial D_{1}}{\partial \tau} \Bigr) D_{21}  +  \fr{\partial D_1}{\partial \tau} \fr{\partial D_{21}}{\partial \tau} \Biggr] \label{delta31} \\
&\times& \Biggl[ - \int d^3 k_{1} \int d^3 k_2 \delta_{\D}(\vec{k}_{12} - \vec{k}) \alpha(\vec{k}_1, \vec{k}_2) \theta_1 (\vec{k}_1) K_{21} (\vec{k}_2) \Biggr] \nonumber \, , \\
&\equiv& \Biggl[\fr{3}{2} \Omega_{m} {\cal H}^2 D_1 D_{21}  +  \fr{\partial D_1}{\partial \tau} \fr{\partial D_{21}}{\partial \tau} \Biggr] K_{31}(\vec{k}) \nonumber \, , \\
\Biggl[ \fr{\partial^2 D_{32}}{\partial \tau^2} + {\cal H} \fr{\partial D_{32}}{\partial \tau} - \fr{3}{2} \Omega_{m} {\cal H}^2 D_{32} \Biggr] K_{32}(\vec{k}) &=& \Biggl[\Bigl(\fr{\partial^2 D_{1}}{\partial \tau^2} + {\cal H} \fr{\partial D_{1}}{\partial \tau} \Bigr) D_{22}  +  \fr{\partial D_1}{\partial \tau} \fr{\partial D_{22}}{\partial \tau} \Biggr] \label{delta32} \\
&\times& \Biggl[ - \int d^3 k_{1} \int d^3 k_2 \delta_{\D}(\vec{k}_{12} - \vec{k}) \alpha(\vec{k}_1, \vec{k}_2) \theta_1 (\vec{k}_1) K_{22} (\vec{k}_2) \Biggr] \nonumber \, , \\
&\equiv& \Biggl[\fr{3}{2} \Omega_{m} {\cal H}^2 D_1 D_{22}  +  \fr{\partial D_1}{\partial \tau} \fr{\partial D_{22}}{\partial \tau} \Biggr] K_{32}(\vec{k}) \nonumber \, , \\
\Biggl[ \fr{\partial^2 D_{33}}{\partial \tau^2} + {\cal H} \fr{\partial D_{33}}{\partial \tau} - \fr{3}{2} \Omega_{m} {\cal H}^2 D_{33} \Biggr] K_{33}(\vec{k}) &=& \Biggl[-\Bigl(\fr{\partial^2 D_{21}}{\partial \tau^2} + {\cal H} \fr{\partial D_{21}}{\partial \tau} \Bigr) D_{1} + \Bigl(\fr{\partial^2 D_{1}}{\partial \tau^2} + {\cal H} \fr{\partial D_{1}}{\partial \tau} \Bigr) D_{1}^2 \nonumber \\ &+& 2 D_1 \Bigl( \fr{\partial D_1}{\partial \tau} \Bigr)^2 - \fr{\partial D_1}{\partial \tau} \fr{\partial D_{21}}{\partial \tau} \Biggr] \label{delta33} \\
&\times& \Biggl[ - \int d^3 k_{1} \int d^3 k_2 \delta_{\D}(\vec{k}_{12} - \vec{k}) \alpha(\vec{k}_1, \vec{k}_2) K_{21} (\vec{k}_1) \delta_1 (\vec{k}_2) \Biggr] \nonumber \, , \\
&\equiv& \Biggl[ -\fr{3}{2} \Omega_{m} {\cal H}^2 D_1 D_{21} - \fr{\partial D_1}{\partial \tau} \fr{\partial D_{21}}{\partial \tau} + D_1 \Bigl( \fr{\partial D_1}{\partial \tau} \Bigr)^2 \Biggr] K_{33}(\vec{k}) \nonumber \, , \\
\Biggl[ \fr{\partial^2 D_{34}}{\partial \tau^2} + {\cal H} \fr{\partial D_{34}}{\partial \tau} - \fr{3}{2} \Omega_{m} {\cal H}^2 D_{34} \Biggr] K_{34}(\vec{k}) &=& \Biggl[-\Bigl(\fr{\partial^2 D_{22}}{\partial \tau^2} + {\cal H} \fr{\partial D_{22}}{\partial \tau} \Bigr) D_{1} - \fr{\partial D_1}{\partial \tau} \fr{\partial D_{22}}{\partial \tau}  \Biggr] \label{delta34} \\
&\times& \Biggl[ - \int d^3 k_{1} \int d^3 k_2 \delta_{\D}(\vec{k}_{12} - \vec{k}) \alpha(\vec{k}_1, \vec{k}_2) B_2 (\vec{k}_1) \delta_1 (\vec{k}_2) \Biggr] \nonumber \, , \\
&\equiv& \Biggl[-\fr{3}{2} \Omega_{m} {\cal H}^2 D_1 D_{22} - \fr{\partial D_1}{\partial \tau} \fr{\partial D_{22}}{\partial \tau}+\fr{1}{2}D_1 \Bigl( \fr{\partial D_1}{\partial \tau} \Bigr)^2    \Biggr] K_{34}(\vec{k}) \nonumber \, , \\
\Biggl[ \fr{\partial^2 D_{35}}{\partial \tau^2} + {\cal H} \fr{\partial D_{35}}{\partial \tau} - \fr{3}{2} \Omega_{m} {\cal H}^2 D_{35} \Biggr] K_{35}(\vec{k}) &=& \fr{1}{2} \Biggl[\fr{\partial D_1}{\partial \tau} \fr{\partial D_{21}}{\partial \tau} -D_1\Bigl(\fr{\partial D_1}{\partial \tau} \Bigr)^2 \Biggr] \label{delta35} \\ &\times& \Biggl[-\int d^3 k_{1} d^3 k_2 \delta_{\D}(\vec{k}_{12}-\vec{k}) \beta(\vec{k}_1, \vec{k}_2) \Bigl(\theta_1 (\vec{k}_1) K_{21} (\vec{k}_2) + K_{21} (\vec{k}_1) \theta_1 (\vec{k}_2) \Bigr) \Biggr] \nonumber \, , \\
\Biggl[ \fr{\partial^2 D_{36}}{\partial \tau^2} + {\cal H} \fr{\partial D_{36}}{\partial \tau} - \fr{3}{2} \Omega_{m} {\cal H}^2 D_{36} \Biggr] K_{36}(\vec{k}) &=& \fr{1}{2} \fr{\partial D_1}{\partial \tau} \fr{\partial D_{22}}{\partial \tau} \label{delta36} \\
&\times& \Biggl[-\int d^3 k_{1} d^3 k_2 \delta_{\D}(\vec{k}_{12} - \vec{k}) \beta(\vec{k}_1, \vec{k}_2) \Bigl(\theta_1 (\vec{k}_1) K_{22} (\vec{k}_2) + K_{22} (\vec{k}_1) \theta_1 (\vec{k}_2) \Bigr) \Biggr] \nonumber \, ,  \ea
One can rewrite the temporal parts of the above Eqs. (\ref{delta31})-(\ref{delta36}) with the proper initial conditions obtained from the EdS solutions at the early epoch to get the fastest growing mode solutions,
\ba \fr{d^2 D_{31}}{d a^2} + \fr{3}{2a} \Bigl(1 - w \Omega_{de} \Bigr) \fr{d D_{31}}{d a} - \fr{3}{2a^2} \Omega_{m} D_{31} &=& \fr{3}{2a^2} \Omega_{m} D_1 D_{21} + \fr{d D_1}{d a} \fr{d D_{21}}{d a} \label{delta312} \\
&{\rm with}& D_{31}(a_i) = 0, \hspace{0.1in} \fr{d D_{31}}{da} \Bigl|_{a_i} = \fr{20}{441} a_i^2 \nonumber \, , \\
\fr{d^2 D_{32}}{d a^2} + \fr{3}{2a} \Bigl(1 - w \Omega_{de} \Bigr) \fr{d D_{32}}{d a} - \fr{3}{2a^2} \Omega_{m} D_{32} &=& \fr{3}{2a^2} \Omega_{m} D_1 D_{22} + \fr{d D_1}{d a} \fr{d D_{22}}{d a} \label{delta322} \\
&{\rm with}& D_{32}(a_i) = 0, \hspace{0.1in} \fr{d D_{32}}{da} \Bigl|_{a_i} = -\fr{4}{441} a_i^2 \nonumber \, , \\
\fr{d^2 D_{33}}{d a^2} + \fr{3}{2a} \Bigl(1 - w \Omega_{de} \Bigr) \fr{d D_{33}}{d a} - \fr{3}{2a^2} \Omega_{m} D_{33} &=& -\fr{3}{2a^2} \Omega_{m} D_1 D_{21} - \fr{d D_1}{d a} \fr{d D_{21}}{d a} + D_1 \Bigl( \fr{d D_1}{d a} \Bigr)^2 \label{delta332} \\
&{\rm with}& D_{33}(a_i) = 0, \hspace{0.1in} \fr{d D_{33}}{da} \Bigl|_{a_i} = \fr{78}{441} a_i^2 \nonumber \, , \\
\fr{d^2 D_{34}}{d a^2} + \fr{3}{2a} \Bigl(1 - w \Omega_{de} \Bigr) \fr{d D_{34}}{d a} - \fr{3}{2a^2} \Omega_{m} D_{34} &=& -\fr{3}{2a^2} \Omega_{m} D_1 D_{22} -\fr{d D_1}{d a} \fr{d D_{22}}{d a} +\fr{1}{2} D_1\Bigl( \fr{d D_1}{d a} \Bigr)^2     \label{delta342} \\
&{\rm with}& D_{34}(a_i) = 0, \hspace{0.1in} \fr{d D_{34}}{da} \Bigl|_{a_i} = \fr{53}{441} a_i^2 \nonumber \, , \\
\fr{d^2 D_{35}}{d a^2} + \fr{3}{2a} \Bigl(1 - w \Omega_{de} \Bigr) \fr{d D_{35}}{d a} - \fr{3}{2a^2} \Omega_{m} D_{35} &=& \fr{1}{2} \Biggl[ -D_1 \Bigl(\fr{d D_1}{d a} \Bigr)^2 + \fr{d D_1}{d a} \fr{d D_{21}}{d a}\Biggr] \label{delta352} \\
&{\rm with}& D_{35}(a_i) = 0, \hspace{0.1in} \fr{d D_{35}}{da} \Bigl|_{a_i} = -\fr{24}{441} a_i^2 \nonumber \, , \\
\fr{d^2 D_{36}}{d a^2} + \fr{3}{2a} \Bigl(1 - w \Omega_{de} \Bigr) \fr{d D_{36}}{d a} - \fr{3}{2a^2} \Omega_{m} D_{36} &=& \fr{1}{2} \fr{d D_1}{d a} \fr{d D_{22}}{d a} \label{delta362} \\
&{\rm with}& D_{36}(a_i) = 0, \hspace{0.1in} \fr{d D_{36}}{da} \Bigl|_{a_i} = -\fr{5}{441} a_i^2 \nonumber \, , \ea
where we use the fastest growing mode solutions for the EdS universe
\ba D_{31}^{(\EdS)} &=& \fr{5}{882} \Bigl(49 a^3 - 90 a_i a^2 + 41 a_i^2 a \Bigr), \,\,\,\,
D_{32}^{(\EdS)} = \fr{-1}{882} \Bigl(49 a^3 - 90 a_i a^2 + 41 a_i^2 a \Bigr) \, , \nonumber \\
D_{33}^{(\EdS)} &=& \fr{-3}{882} \Bigl(49 a^3 - 45 a_i a^2 - 4 a_i^2 a \Bigr), \,\,\,\,
D_{34}^{(\EdS)} = \fr{2}{882} \Bigl(49 a^3 - 90 a_i a^2 + 41 a_i^2 a \Bigr) \, , \label{D316EdS} \\
D_{35}^{(\EdS)} &=& \fr{3}{882} \Bigl(7 a^3 - 30 a_i a^2 + 23 a_i^2 a \Bigr), \,\,\,\,
D_{36}^{(\EdS)} = \fr{-2}{882} \Bigl(7 a^3 - 9 a_i a^2 + 2 a_i^2 a \Bigr) \, . \nonumber \ea
In the above equation (\ref{D316EdS}), we use the initial Gaussinity condition of $\delta^{(3)}$ ({\it i.e.} $D_{3i}(a_i) = 0$) to obtain the coefficients for the last terms of $D_{3i}$.

One can find the third order kernels ($F_{3i}(\kk)$) from the above Eqs. (\ref{delta31})-(\ref{delta36}). For example, one obtain $F_{31}^{(s)}$ as
\ba K_{31}(\vec{k}) &=& - \int d^3 k_{1} \int d^3 k_2 \delta_{\D}(\vec{k}_{12} - \vec{k}) \alpha(\vec{k}_1, \vec{k}_2) \theta_1 (\vec{k}_1) K_{21} (\vec{k}_2) \nonumber \\
&=&  \int d^3 q_1 d^3 q_2 d^3 q_{3}  \delta_{\D}(\vec{q}_{1} +\vec{q}_2+\vec{q}_3 - \vec{k}) \alpha(\vec{q}_3, \vec{k}-\vec{q}_3) \alpha(\vec{q}_1, \vec{q}_2) \delta_1(\vec{q}_1) \delta_1(\vec{q}_2) \delta_1 (\vec{q}_3) \, , \nonumber \\
F_{31}(a,\vec{q}_1,\vec{q}_2,\vec{q}_3) &\equiv& c_{31}(a) \alpha(\vec{q}_3, \vec{k}-\vec{q}_3) \alpha(\vec{q}_1, \vec{q}_2), \,\,\,
F_{31}^{(s)}(a,\vec{q}_1, \vec{q}_2, \vec{q}_3) = \fr{c_{31}}{3!} \Biggl[F_{31}(a, \vec{q}_1, \vec{q}_2, \vec{q}_3) + {\rm perm} \Biggr] \nonumber \, , \\
F_{31}^{(s)}(a,\vec{q},-\vec{q},\vec{k}) &=& \fr{c_{31}}{3!} \Biggl[-2 x^2 \Bigl(\fr{1+r^2}{r^2} \Bigr) \Biggr] \label{F31s} \, . \ea
One can repeat the above process to obtain
\ba F_{32}(a, \vec{q}_1, \vec{q}_2, \vec{q}_3) &=& -c_{32} \alpha(\vec{q}_3, \vec{k}-\vec{q}_3) \beta(\vec{q}_1, \vec{q}_2) \, , \nonumber \\ F_{32}^{(s)}(a, \vec{q}, -\vec{q}, \vec{k}) &=& \fr{c_{32}}{3!} 4x^2 \Bigl(\fr{1+r^2}{r^2} \Bigr) \label{F32s} \, , \\
F_{33}(a,\vec{q}_1, \vec{q}_2, \vec{q}_3) &=& -c_{33} \alpha(\vec{k}-\vec{q}_3,\vec{q}_3) \alpha(\vec{q}_1, \vec{q}_2) \, ,\nonumber \\
F_{33}^{(s)}(a,\vec{q}, -\vec{q}, \vec{k}) &=& \fr{c_{33}}{3!} \fr{-4(1+r^2)+2(1+4r^2-r^4)x^2}{(1+r^2+2rx)(1+r^2-2rx)}  \label{F33s} \, , \\
F_{34}(a,\vec{q}_1, \vec{q}_2, \vec{q}_3) &=& c_{34} \alpha(\vec{k}-\vec{q}_3,\vec{q}_3) \beta(\vec{q}_1, \vec{q}_2) \, , \nonumber \\
F_{34}^{(s)}(a,\vec{q}, -\vec{q}, \vec{k}) &=& \fr{c_{34}}{3!} 4x^2 \label{F34s} \, , \\
F_{35}(a,\vec{q}_1, \vec{q}_2, \vec{q}_3) &=& 2c_{35} \beta(\vec{k}-\vec{q}_3,\vec{q}_3) \alpha(\vec{q}_1, \vec{q}_2) \, ,  \nonumber \\
F_{35}^{(s)}(a, \vec{q}, -\vec{q}, \vec{k}) &=& \fr{c_{35}}{3!} \fr{-8r^2(1+r^2) + 4(-1+4r^2+r^4)x^2}{r^2(1+r^2+2rx)(1+r^2-2rx)} \label{F35s} \, , \\
F_{36}(a,\vec{q}_1, \vec{q}_2, \vec{q}_3) &=& -2c_{36} \beta(\vec{k}-\vec{q}_3,\vec{q}_3) \beta(\vec{q}_1, \vec{q}_2) \, , \nonumber \\
F_{36}^{(s)}(a,\vec{q}, -\vec{q}, \vec{k}) &=& \fr{c_{36}}{3!} \fr{ 8x^2}{r^2} \label{F36s} \ea

One can obtain $c_{3i}(a)$ numerically at any epoch by solving the above Eqs.(\ref{delta312})-(\ref{delta362}).
In EdS case, $c_{3i}$ have been known as
\be c_{31} = \fr{5}{18} \,\, , c_{32} = -\fr{1}{18} \,\, , c_{33} = -\fr{1}{6} \,\, , c_{34} = \fr{1}{9} \,\, , c_{35} = \fr{1}{42} \,\, , c_{36} = -\fr{1}{63} \label{c3iEdS} \, . \ee
However, the above values are not exact because they are not the coefficients for the fastest growing mode solutions as shown in Eq. (\ref{D316EdS}). The above values given in Eq.(\ref{c3iEdS}) used in the kernels in the reference \cite{08061437}. For the second order, this is a good approximation but not for the third order.

Now one can explicitly write $\hat{\delta}^{(3)}(a,\vec{k})$ as
\be \hat{\delta}^{(3)}(a,\vec{k}) \equiv D_{1}^3(a) \sum_{i=1}^{6} c_{3i}(a) K_{3i}(\vec{k})= D_{1}^3(a) \Biggl[ c_{31}(a) K_{31} (\vec{k}) + \cdots + c_{36}(a) K_{36}(\vec{k}) \Biggr] \label{delta32} \ee
Thus, one can calculate $P_{22}(a,k)$ and $P_{13}(a,k)$ at any epoch.
\ba P_{22}(a,k) &=& D_{1}^4(a) 2 \int d^3 q \Bigl[ F_{2}^{(s)}(\vec{k}-\vec{q},\vec{q}) \Bigr]^2 P_{11}(k-q) P_{11}(q) \nonumber \\ &=& D_{1}^4(a) 2 \int (2 \pi) q^2 d q \int_{-1}^{1} dx \Bigl[ F_{2}^{(s)}(\vec{k}-\vec{q},\vec{q}) \Bigr]^2 P_{11}(kr) P_{11}(k\sqrt{1+r^2-2rx}) \nonumber \\ &=& \fr{(2\pi)^{-2} k^3}{2} \int_{0}^{\infty} dr P_{11}(kr) \int_{-1}^{1} dx P_{11}(k\sqrt{1+r^2-2rx}) \nonumber \\ &\times& \Biggl[\fr{(c_{21} + 2c_{22})r + (c_{21} - 2c_{22})x - 2c_{21} rx^2 }{(1+r^2-2rx)} \Biggr]^2 \label{P22new2} \, , \\
P_{22}^{({\rm EdS})}(a,k) &=& D_{1}^4(a) \fr{(2\pi)^{-2} k^3}{2} \int_{0}^{\infty} dr P_{11}(kr) \int_{-1}^{1} dx P_{11}(k\sqrt{1+r^2-2rx}) \Biggl[\fr{3r + x - 10 rx^2 }{7(1+r^2-2rx)} \Biggr]^2 \label{P22new2Eds} \, , \\
P_{13}(a,k) &=& 6 D_{1}^4(a) P_{11}(k) \int d^3 q P_{11}(q) \Bigl[ F_{3}^{(s)}(a,\vec{q},-\vec{q},\vec{k}) \Bigr] \label{P13new2} \\ &=& D_{1}^4 (2\pi)^{-2} k^3 P_{11}(k) \int_{0}^{\infty} dr P_{11}(kr) \Biggl[ 2c_{35} r^{-2} -\fr{1}{3} \Bigl( 4 c_{31} -8 c_{32} +3c_{33} +24c_{35} - 16 c_{36} \Bigr) \nonumber \\ &-& \fr{1}{3}\Bigl(4 c_{31} -8c_{32} +12c_{33}-8c_{34}+6c_{35} \Bigr)r^2 + c_{33} r^4 + \Bigl(\fr{r^2-1}{r} \Bigr)^3 \ln \Bigl|\fr{1+r}{1-r} \Bigr| \Bigl(c_{35} - \fr{1}{2}c_{33}r^2 \Bigr) \Biggr] \nonumber \, , \\
P_{13}^{({\rm EdS})}(a,k) &=& D_{1}^4(a) (2\pi)^{-2} k^3 P_{11}(k) \int_{0}^{\infty} dr P_{11}(kr) \Biggl[ \fr{1}{21} r^{-2} -\fr{79}{126} + \fr{25}{63}r^2 -\fr{1}{6}r^4 \nonumber \\ &+& \Bigl(\fr{r^2-1}{r} \Bigr)^3 \ln \Bigl|\fr{1+r}{1-r} \Bigr| \Bigl( \fr{1}{42}+\fr{1}{12}r^2 \Bigr) \Biggr] \label{P13new2EdS} \, , \ea
where we use
\be |\vec{q}| = r |\vec{k}|, \,\, \vec{q} \cdot \vec{k} = x |\vec{q}| |\vec{k}|, \,\, k_1^2 = |\vec{q}|^2 = r^2 k^2, \,\, k_2^2 = |\vec{k}-\vec{q}|^2 = k^2 + q^2 -2 \vec{k} \cdot \vec{q} = k^2 ( 1+ r^2 -2rx), \,\, \vec{k}_1 \cdot \vec{k}_2 = k^2 r (x-r) \label{qk} \ee
We also check the dependence of $c_{3i}$ on $a_i$. When we compare $a_i=\fr{1}{50}$ to $a_i = \fr{1}{1000}$, there are sub percent level differences.


\begin{thebibliography}{99}

\bibitem{0112551} F.~Bernardeau, S.~Colombi, E.~Gaztanaga, and R.~Scoccimarro, Phys.\ Rept.\ {\bf 367}, 1 (2002) [arXiv:astro-ph/0112551].

\bibitem{13112724} F.~Bernardeau, Les Houches Summer School 'Post-Planck Cosmology' [arXiv:1311.2724].

\bibitem{Goroff} M.~H.~Goroff, B.~Grinstein, S.-J.~Rey, and M.~B.~Wise, Astrophys.\  J. {\bf 311}, 6 (1986).

\bibitem{Makino} N.~Makino, M.~Sasaki, and Y.~Suto, Phys.\ Rev.\ D {\bf 46}, 585 (1992).

\bibitem{9807211} M.~Kamionkowski and A.~Buchalter, Astrophys.\ J. {\bf 514}, 7 (1999) [arXiv:astro-ph/9807211].

\bibitem{08061437} R.~Takahashi, Prog.\ Theor.\ Phys. {\bf 120}, 549 (2008) [arXiv:0806.1437].

\bibitem{0509418} M.~Crocce and R.~Scoccimarro, Phys.\ Rev.\ D {\bf 73}, 063519 (2006) [arXiv:astro-ph/0509418].

\bibitem{0509419} M.~Crocce and R.~Scoccimarro, Phys.\ Rev.\ D {\bf 73}, 063520 (2006) [arXiv:astro-ph/0509419].

\bibitem{12034260} C.~Rampf and T.~Buchert, JCAP {\bf 1206}, 021 (2012) [arXiv:1203.4260].

\bibitem{14012226} S.~Lee, Phys.\ Rev.\ D {\bf 89}, 084017 (2014) [arXiv:1401.2226].

\bibitem{14043813} S.~Lee, [arXiv:1404.3813].



\bibitem{camb} A.~Lewis and A.~Challinor, URL: http://www.camb.info



\bibitem{9408025} C.~Wetterich, Astron.\ Astrophys. {\bf 301}, 321 (1995) [arXiv:hep-th/9408025].

\bibitem{0012139} M.~Doran, M.~Lilley, J.~Schwindt, and C.~Wetterich, Astrophys.\ J. {\bf 559}, 501 (2001) [arXiv:astro-ph/0012139].

\bibitem{0504650} S.~Lee, Phys.\ Rev.\ D {\bf 71}, 123528 (2005) [arXiv:astro-ph/0504650].



\end{thebibliography}
\end{document}